# The Effects of Orographic Geometry on Supercell Thunderstorms

Galen Smith[1] and Yuh-Lang Lin[1,2,@]
[1]Department of Energy and Environmental Systems
[2]Department of Physics
North Carolina A&T State University

[@]Corresponding Author Address: Dr. Yuh-Lang Lin, 302H Gibbs Hall, EES, North Carolina A&T State University, 1601 E. Market St., Greensboro, NC 27411.
Email: ylin@ncat.edu. Web: http://mesolab.ncat.edu

**Abstract**
The effect of elongated bell-shaped mountain orientation on supercell thunderstorms is numerically investigated using the Bryan Cloud Model 1 (CM1). The orography is varied by three mountain heights and is varied in four different positions, effectively producing 12 different terrain configurations. It is found that the different orientations produce variations in the supercell life cycle with shorter cycles for higher inflow rates. Furthermore, these cycles are associated with the storm reaching its minimum intensity just after a peak rain period. Moreover, the effect of stronger inflow was seen before direct storm-terrain interactions started. The higher inflow also played a significant role in increasing rainfall rate and areal extent, to the point that further convection, associated with the cold pool, was triggered adding to rainfall amount. Using a stricter form of the National Weather Service Tornado Detection Algorithm to investigate the tornadic nature of simulated supercells; it is found that terrain blocking effects are dominative and that elongating the terrain axis approximately parallel to the propagation vector produced the strongest potential to generate a tornadic supercell thunderstorm. Although the simulated case with the highest mountain produced the most tornadic thunderstorms it is seen that increasing the terrain height alone is not sufficient to make tornadogenesis more probable as more tornadic supercells were simulated with lower heights compared to moderate terrain heights.

## 1. Introduction

Tornadic supercell thunderstorms are known to primarily occur in the Midwestern United States with the common pattern of the coupling of the updraft region of the polar and sub-tropical jets and a low-level jet transporting warm moist air near this updraft region. This sets a stage that is perfect for supercellular convection and tornadogenesis through copious amounts of CAPE and a veering wind profile. Broyles and Crosbie (2004) looked at the long track climatology of violent tornadoes and identified several smaller tornado alleys. These localized tornado alleys show that despite the belief that tornado alley is "flat" there are in fact significant elevation changes, several hundred meters in many cases (Fig. 1). Although Broyles and Crosbie's investigation is climatological, it points us into the direction to investigate if terrain can induce tornadogenesis or if the terrain is coincidental to the stronger influence of jets and CAPE. Furthermore, if we look at the tracks of the tornadoes in these sub-regions, it appears that there is not a general "approach angle" of the storm on the terrain (Fig. 2).

Many case studies have been done on tornadoes that have occurred in complicated terrain (e.g. Bluestein, 2000; Homar, 2003; LaPenta et al., 2005; Bosart et al., 2006; Schneider, 2009). These studies have generally found that the terrain modified the environment to be more favorable for tornadogenesis. Furthermore, a few researchers have directly attributed the formation of a tornado to terrain effects (Bosart, et al., 2006; LaPenta et al., 2005). In addition to these case studies, Homar et al. (2003) conducted a numerical investigation, using the 5$^{th}$ generation of the Pennsylvania State University - National Center for Atmospheric Research Mesoscale Model (MM5v3, Dudhia, 1993; Grell et al., 1995), on a tornado that occurred in Spain on 28 AUG 1999. They specifically investigated how the terrain resolution affected the generation of rotation in the supercell, although their study did not attempt to reproduce the tornado. They found that the rotation could be attributed to large scale features (20-50 km) and that the storm's intensification could be attributed to small scale terrain features (2-5 km). Additionally, Ćurić et al. (2007) investigated vertical vorticity characteristics of individual cumulonimbus clouds using the Advanced Regional Prediction System (ARPS, Xue et al., 2001). They concluded that vertical vorticities were enhanced at the lowest levels of the simulated clouds when terrain was included in the simulation.

Although the above investigations have shown a positive correlation between vertical rotation enhancement and terrain, they are of actual tornadic events and use the observed data to initialize the simulations. This may somewhat cloud the actual connection as observations have inherent uncertainties and/or limitations additionally the observations are being modified to suit the modified terrain. It is known that tornadoes can occur in environments with quite complicated terrain which is difficult to investigate. To counter this uncertainty some idealized studies using idealized terrain have been done to investigate the effects that orography has on supercell thunderstorms (Markowski and Dotzek, 2011; Smith 2014).

MD11 investigated how the environmental modifications of a bell-shaped mountain and a flat toped hill with and incised gap, both 500 m tall with a 10 km half-width, affected supercell thunderstorms generated using the Wiseman and Klemp (1982 – WK82 hereafter) analytical sounding. They aimed the storm such that it would intersect the location of the maximum positive vertical vorticity anomaly on the lee side of the bell-shaped mountain, as identified by an environmental simulation in which a storm was not initiated. They determined that the most notable change in the supercells evolution is a gradual strengthening of the midlevel and low-level updraft by upslope winds; then followed by a weakening of the updrafts but a rapid spin-up of low-level vorticity when the storm propagates over the vertical vorticity anomaly, even if the storms updraft is being weakened by a region of increased convective inhibition.

S14 investigated how bell-shaped mountain blocking affects the development, structure, intensity, and track of supercell thunderstorms. They used the same model resolution and sounding as MD11, but used calm surface winds. Their findings suggested that the track of the supercell thunderstorm was shifted towards the left of the storms motion slightly (more so for higher mountains). They also noted that the structure of the storm was modified such that the main updraft was larger and stronger as a result of the terrain diverting more air for the storm to ingest. The additional airstream also created a larger anvil cloud and increased the areal extent of the rainfall. Additionally, S14 found that as the storm approached the mountain peak terrain blocking effects allowed the generation of vertical vorticity along the gust front and was intensified by the main updraft. They stated that these vortices most likely would have been tornadoes if the resolution of their simulations were higher.

Although the terrain blocking effects of bell-shaped mountains has been investigated by S14, there is additional work to be done such as using an elongated mountain and modifying the approach angle of the storm to determine if there is a favorable configuration that would be more likely to enhance tornadogenesis potential. Therefore, in this investigation, we will simplify the terrain to bell-shaped mountains that have been elongated and have different orientations to vary the direction of the mean wind in relation to the terrain while preserving the storms motion. Along with varying the orientation the terrain heights will be varied (500, 1000, and 1500m), which give corresponding unsaturated moist Froude numbers ($F_w$) of; 1.78, 0.89, and 0.59. The $F_w$ is defined as $F_w = U/(N_w h)$ (e.g., see Lin 2007, Chen and Lin 2005, and Emanuel 1994), where $U$ is the basic wind, $N_w$ is the unsaturated moist Brunt–Väisälä frequency, and $h$ is the mountain height. Both the basic wind and the Brunt–Väisälä frequency are averaged over the depth of the mountain.

## 2. Methodology
### 2.1. Model Selection and Description

Our simulations utilize the Bryan Cloud model version 1, Release 16 (CM1) (Bryan and Fritsch 2002). CM1 is a non-hydrostatic idealized numerical model designed to utilize high resolutions, particularly for severe local storms which contain deep moist convection. The governing equations that CM1 utilizes conserve mass and total energy, but they are not fully conserved in the model due to limitations in numerical integration. The CM1 introduced new equations for calculating gradients that better conserve mass and energy in simulations containing terrain and that employ stretched vertical coordinate. CM1 uses the Gal-Chen and Somerville (1975) terrain-following coordinates to map the model levels to the terrain while the model top is at constant height, and the governing equations are adapted from those described by Wicker and Skamarock (2002). The advection terms are discretized using fifth-order spatial finite difference and artificial diffusion, which may be applied both horizontally and vertically using separate coefficients. The sub-grid turbulence parameterization is similar to the parameterization of Deardorff (1980). CM1 has several options in microphysics parameterization schemes.

### 2.2. Model Configuration and Experimental Design

The domain is 300 × 100 × 18 km in the x, y, and z directions, respectively. In order to study the impacts of terrain a storm, the grid is stationary, instead of moving with the storm; otherwise the path of the storm will be affected. The horizontal grid spacing is 500 m; the vertical grid spacing varies from 25 m in the lowest 500 m, to constant 500 m from 11 km to 18 km (74 vertical levels total). The environments are horizontally homogeneous at the start of the simulations except in cases where storms were initialized with a warm bubble, 2 K warmer than the environment, centered at (x, y, z) = (46, 35, 1.4 km) with the origin located at the southwest corner of the domain. The warm bubble depth was 1.4 km and had a horizontal radius of 10 km. Simulations were run for a period of 4 h, and model outputs were every 15 min.

Simulations with storms were initialized in a way that the supercell would arrive near the terrain's peak at approximately the 180[th] min of simulation time when the supercell would be quasi-steady when it interacted with the terrain. In addition to ensuring that the storm would be quasi-steady the location was chosen such that the supercell would propagate as close to the peak of the terrain as possible.

The terrain used in this research is centered at (x, y) = (200, 50 km) with the origin located at the southwest corner of the domain and is an elongated bell-shaped mountain with half-widths of $a = 10$ km and $b = 20$ km along the semi-minor and semi-major axis and the height is varied from flat terrain to the highest peak $h_m = 500$ m, 1000 m, and 1500 m. The elongated mountain is rotated to vary the direction of the mean wind and the amount of blocking induced by the terrain. The bell-shaped mountain is rotated by applying the standard rotation operator and is defined by the following:

$$h(x,y) = \frac{h_m}{\left[1 + \left(\frac{(x-x_0)\cos\theta - (y-y_0)\sin\theta}{a}\right)^2 + \left(\frac{(x-x_0)\sin\theta + (y-y_0)\cos\theta}{b}\right)^2\right]^{3/2}}$$

where $h_m$ is the mountain height, $a$ and $b$ are the mountain half-widths, and $(x_0, y_0)$ is the center of the mountain and $\theta$ is the angle about which the terrain has been rotated.

The lower boundary is free-slip and the upper boundary utilizes a Rayleigh damping layer (Durran and Klemp, 1983) in the uppermost 3 km of the model domain in order to allow the gravity waves generated by the terrain and convection to propagate upward out of the domain. In addition, lateral boundaries are open and radiative. Surface heat fluxes, atmospheric radiative heating, and the Coriolis force are set to zero for our simulations. The simulation uses the NASA-Goddard version of the Lin-Farley-Orville (LFO) microphysics parameterization scheme (Lin et al., 1983).

The environments of the simulated storms are initialized with a sounding very similar to the analytic sounding of WK82 (Fig. 3) and a warm bubble as descried above. Although it has been found that models initialized with the WK82 (standard) sounding resulted in a moist absolutely unstable layer when ascending over a relatively small hill (Bryan and Fritsch, 2000; MD11); we believe that this was due to issues with the way previous models handled momentum and energy, as our model output did not indicate a moist absolutely unstable layer generated by the terrain.

The sounding has a mixed layer convective available potential energy (MLCAPE) of 1955 J kg$^{-1}$ and a mixed layer convective inhibition (MLCIN) of 33 J kg$^{-1}$. The environmental wind profile is defined by the analytical quarter-circle hodograph described by WK82 (Fig. 3). The WK82 wind profile has a bulk shear (0–6 km shear vector magnitude) of 32 m s$^{-1}$ and storm-relative helicity (SRH) of 191 m$^2$s$^{-2}$. This wind profile gives a value of approximately 15 for the *supercell composite parameter* (Thompson, et al., 2005, 2007) not surprising since the vertical moisture and wind profile for the WK82 was developed to simulate supercellular convection (although many of the included soundings in their analysis were tornadic). Moreover, the *significant tornado parameter* (STP) (Thompson, et al., 2005, 2007) is greater than 2. Note that values of the STP greater than 1 are associated with the majority of tornadoes stronger than F2 while non-tornadic supercells are associated with STP values less than 1 (Thompson, et al., 2005, 2007).

*2.3 Modified Detection Algorithm for Tornado/Tornadogenesis*
In order to declare a storm tornadic we must develop a method for declaring what exactly we shall call a tornado. We will determine if a storm is tornadic in a manner quite similar to the tornado detection algorithm that is used by the National Weather Service to detect tornado signatures in radar observations. The tornado detection algorithm is modified and our method for tornado declaration is as follows:
1. A vortex is present at the lowest model level (LML, 12.5 m AGL) and the cross vortex shear is greater than 25 m s$^{-1}$ at the LML, and the vortex has depth of at least 1.5 km

or

2. A vortex is present at the LML, the vortex has a depth of at least 1.5 km, and the cross vortex shear is greater than 36 m s$^{-1}$ anywhere in the lowest 500 m of the vortex.

Any storm that meets these qualifications we shall deem as a tornadic supercell for our investigation. Figure 4 shows an example of the difference between a flow field that comes from a simulation that meet criteria 1 and another simulation that meets everything in criteria 1 except for having a vortex. If we assume that a vortex requires at least 4 grid points to be adequately resolved; then the minimum vorticity that could meet either of these requirements is 0.0225 s$^{-1}$.

## 3. Terrain induced environmental modifications
*3.1 Environmental Simulation (Mountain Only – MTNO)*
(a) Bell-Shaped Mountain

S14 investigated the environmental evolution of the WK82 sounding in response to idealized bell-shaped mountains of 500, 1000, and 1500 m heights. Their analysis showed that there was reduced MLCIN over and around the underlying terrain (mostly associated with a reduction in the distance from the surface to the Lifting Condensation Level (LCL). Additionally, the presence of gravity waves modified MLCAPE and MLCIN, which fed back to these waves and made them stronger for each progressive height. Furthermore, the waves for the 1000 and 1500 m simulations were strong enough to initiate supercellular convection approximately 50 km north-east of the terrain peak. Finally, their analysis of the low-level vorticity and wind field show that the 1500 m terrain simulation is the only one that generating a closed pair of counter-rotating vortices.

(b) Environment of Mountains with Modified Geometry

Simulations were performed with mountains of 500, 1000, and 1500 m heights and the mountains were varied through four basic positions such that the semi-major axis is nearly perpendicular to the layer averaged winds (RM45), then nearly parallel (RP45), then two configurations that are roughly 45 deg toward and away (cases 2A, 2B). Rotating the terrain allowed the blocking effect of the terrain to be varied, from strongest (case RM45) to weakest (case RP45). The effective $F_w$ are given in (Table 1), which is determined as follows:

$$F_{we} = F_w \left[ \frac{I - Abs(Max(U \cdot T_n, C_s \cdot T_n))}{I} \right]$$

where $F_{we}$ is the effective $F_w$, $I$ is the interval size defined as: $I \equiv \sum Abs[Max(U \cdot T_n, C_s \cdot T_n)]]$, $U$ is the layer averaged basic wind, $C_s$ is the storm motion vector, and $T_n$ is the terrain unit vector normal to the semi-minor axis.

The environmental modifications induced by the terrain were, as expected, similar to that of the idealized bell-shaped mountains in S14. There was a general region over the terrain where MLCIN (MLCAPE) was reduced (increased) because of a reduction the Lifting Condensation Level (LCL; Fig. 5). These simulations showed alternating increases and decreases in MLCAPE and MLCIN, associated with gravity waves, are seen clearly by the end of the first hour. Vertical vorticity (not shown) indicates the existence of convective rolls over the mountain peak in the simulations with the strongest blocking (Figs. 5i-l). Closed wake vortices are only seen in the two simulations with the strongest blocking (Figs. 5i and k).

As with the BSM of S14, supercellular convection was triggered downwind of the mountain in all simulations with mountain heights of 1000 and 1500 m. Interestingly, the terrain configuration (case RM45) that produced the strongest blocking initiated the convection later that of the other terrain configurations. This leads to a general region of reduced CAPE and increased CIN associated with the cold pool outflow from these storms. The cold pools generated in these simulations do not affect our analysis as our focus is on the windward side of the mountains.

*3.2 No Mountain Control Simulation (NMTN)*

To establish a baseline of how CM1 simulates a supercell thunderstorm, we performed a simulation without terrain. This simulation was initiated by the same warm bubble as those for the mountain (MTN) cases, to be presented later in Sec. 4. This will also allow us to better isolate the effects of terrain.

The NMTN simulation produced well-defined right-moving supercell thunderstorms with sustained midlevel (5 km AGL) updrafts (downdrafts) exceeding 30 m s$^{-1}$ (15 m s$^{-1}$) and low-level (500 m AGL) updrafts (downdrafts) exceeding 5 m s$^{-1}$ (10 m s$^{-1}$). Organization of midlevel rotation ($\zeta > 0.003$ s$^{-1}$, where $\zeta$ is the vertical vorticity) was incipient within 30 min of simulation time and was well-organized by 45 min (with $\zeta > 0.02$ s$^{-1}$). Midlevel mesocyclones were sustained throughout the end of the simulations (although cyclic in strength); reflectivity, exhibiting clear indication of the classic supercell structure, from 105 min (Fig. 6). This figure also shows that the midlevel rotation is aligned with the updraft, indicated by rotating winds aligned with the weak echo region, which could make the storm more conducive to tornadogenesis. The vertical vorticities simulated in the control case ranged from 0.02-0.05 s$^{-1}$ at 500 m AGL, from the 60 min out of 180 min. This storm propagates eastward at approximately 15 ms$^{-1}$, with small north to south variation is the location of the storm staying within approximately ±5 km from the location of the warm bubble used to initiate convection.

The NMTN simulation also shows an anticyclonic left-moving storm that propagated out of the domain by hour 2 of simulation time.

## 4. Simulations with Mountains (MTN)

*4.1 Bell-Shaped Mountain*

S14 investigated the effects of bell-shaped mountains with heights of 500, 1000, and 1500 m this effectively varied the amount of terrain blocking, but precluded the ability to investigate the approach angle. S14 showed that even relatively modest terrain could enhance near surface vertical rotation, enhance the intensity of supercellular convection (by diverting additional air into the storm), shift the track of the storm (to the left of the storms motion), and altered the structure of the storm. An additional finding was that the amount of precipitation was increased in the areal extent.

*4.2 Effects of Terrain Geometry on Supercell Thunderstorms*

(a) Orographic effects on Supercell Structure and Development

The structure and development of these storms is quite similar, the storms exhibit nearly identical maturing processes and exhibit the structure of the classical High Precipitation Supercell conceptual model (Lemon and Doswell, 1979) by the 105 min. The storms in each of the simulations remain structurally similar until interacting with the terrain directly. Our analysis is started at the 165 and is stopped at the 180 min. This is the last point at which 1000 and 1500 m terrain cases do not interact with the cold pool on the lee side of the mountain. For the most meaningful interpretation comparisons will be grouped by height 500, 1000, and 1500 m terrain heights; M500, M1000, and M1500 respectively. After discussion with the grouped heights some comments will be made about the overall comparisons between the simulations as a whole.

Starting with M500 an immediately noticeable difference between these simulations is that the cloud updraft area in the M500-2A simulation is nearly half the size of the other terrain orientations (Fig. 7). In addition to the M500-2A simulation having the narrowest updraft region it also has the strongest gust front. Incidentally this is also the point with the highest precipitation rate (nearly 14 cm h$^{-1}$) in M500 (Fig. 13c). As the gust front of the M500-2A simulation starts to weaken the cloud updraft area begins to widen (Figs. 7c and 8c), partially indicative of the cyclic nature of supercell thunderstorms (Burgess et al. 1982; Beck et al. 2006, Klemp and Rotunno 1983; Wicker and Wilhelmson 1995). As the rainfall strengthens in M500-RM45 a stronger gust front is seen developing below the leading edge of the updraft region. The cloud base is the lowest in M500-RP45 and 2A (Fig. 7). Additionally, as these storms propagate towards the terrain, a downdraft is present that advects reflective hydrometeors towards the ground that produces strong reflectivity near the ground (Fig. 8).

The 1000 m terrain simulations (M1000) RP45, 2A, and 2B simulations at the 165 min are close to each other in the supercell cycle despite looking quite different, which can be seen by comparing the difference in appearance between Figs. 9 and 10, especially Figs. 9b and 10d. Although close in the supercells cycle there are differences, such as, the general orientation of the updrafts in these simulations is fairly vertical at the 165 min (Fig. 9). The M1000-RM45 and 2A storms generate a noticeable eastward tilt this leads to a smaller upwind outflow cloud (Fig. 10). The RP45, 2A, and 2B simulations updraft broadens considerably at the mid and upper levels (Fig. 10). We also see that the low-level updraft of M1000-2B narrows considerably (Fig. 10d) as the storm goes through a period of very strong rain (Fig. 13). As the M1000 storms propagate towards the terrain a downdraft advects reflective hydrometeors towards the ground that produces strong reflectivity near the surface (Fig. 10).

The 1500 m simulations (M1500) by far had the strongest cold pools which are associated with very high precipitation (Fig. 13). Moreover, these strong cold pools were able to initiate convection. This is immediately noticeable by the area of updraft several kilometers to the west of the main updraft of the supercell thunderstorm (Figs. 11a and 11d). This initiated convection may also be a factor of blocking increasing the cold pool depth by stronger down slope winds, as there is less indication of convective initiation when there is less blocking (Figs. 11b and 12b). Although this increases the total amount of precipitation it also weakens the storm considerably (Fig. 12d). As the storms propagate over the terrain, the cold pool is blocked by the terrain and the storms are deprived of the additional air lifted by the cold pool and there is a general reduction of the storms updraft as less air is ingested by the storm.

Overall comparison shows, as expected, that M1500 induced more rain both in areal extent and rain rate (Figure 13). Furthermore, there is the least variability from storm to storm in M500 even when propagating over the mountain. Higher terrain height generally, but not always, produced more rain, as shown in the comparison of M500 and M1000. The higher rain rates also deepened cold pools that in M1500 initiated additional convection and producing more rain.

(b) Orographic Effects on Supercell Intensity and Tornadogenesis

The intensity of the storms is investigated by looking at the 1 and 5 km updraft strength and the near surface vorticity. Furthermore, as discussed in the methodology section we will determine if a storm is tornadic to determine which terrain configuration is most favorable for tornadogenesis. Our analysis will focus on the period from the 165 to 210 min.

The storms in all simulations behaved nearly identical throughout the first two hours with minor variations to the storm location and exhibited cyclic intensification and decay, consistent with observations (Burgess et al. 1982; Beck et al. 2006) and previous numerical simulations (Klemp and Rotunno 1983; Wicker and Wilhelmson 1995). Terrain also induced slight changes to the period of the storms cycle with intensity peaks nearly the same as a control simulation without terrain (~75 min) for cases with weak blocking and intensity peaks being closer together for cases with strong blocking. The reduction of the cycle appears to be related to additional air directed into the storm by the terrain. The additional air leads to increased rain and a stronger cold pool that helps to reduce the distance between the rear flank down draft and the storms main updraft weakening the storm until it propagates away allowing the storm re-intensify.

At the 165 min the 1 km AGL updraft of M500-2A, 2B, RP45, and RM45 was 15, 14, 12 and 12 m s$^{-1}$, respectively (Table 2). The downdrafts at this level were 16, 14, 12 and 15 m s$^{-1}$. The updrafts of M500-2A and 2B weaken slightly by the 180 min most likely due to precipitation loading effects, whereas M500-RP45 and RM45 remain about the same magnitude. The updrafts of M500-2A, 2B, and RP45 continue to weaken as the storm propagates over to the lee side of the mountain. The 5 km AGL updrafts and downdrafts are much less affected by the terrain the 2A, 2B, and RP45 cases intensify slightly giving rise to a slight midlevel stretching. The RM45 simulation updraft weakened considerably with its initial interaction with the terrain then the storm nearly recovered to its pre-interaction strength. Terrain blocking effects actually serve to lower the low-level vorticity in the M500-2B and RP45 simulations (Table 2). Although surface vorticity is increased greatly in the M500-RM45 simulation a vortex never formed and did not meet either of our criteria for being declared tornadic. Even though weakening from the 165 to 180 min the M500-2A and RP45 simulations met both criteria to be deemed a tornadic supercell at the 180 min (Table 3).

At the 165 min the 1 km AGL updraft of M1000-2A, 2B, RP45, and RM45 was 13, 12, 12 and 12 m s$^{-1}$, respectively (Table 2). The downdrafts at this level were 16, 15, 14 and 11 m s$^{-1}$. There is a general strengthening of the low-level updraft as the M1000-2A, 2B, and RM45 approach the terrain and couple with upslope winds. The midlevel 5 km AGL updrafts weaken slightly leading to a slight broadening of the wind field as a light dynamic high forms. The low-level updraft weakens as the storm propagates over to the lee side of the mountain as the midlevel updrafts gradually strengthen. As M1000-2A and RP45 approach the terrain the surface vorticity increases slightly and as was with M500-RM45 the M1000-RM45 increased considerably (Table 2). M1000-2B vorticity weakened. M1000-2B, RP45, and RM45 had enough vorticity to cross our minimum vorticity threshold of 0.0225 s$^{-1}$, the vortex was not formed at the LML. M1000-2A simulation met both criteria and is declared tornadic (Table 3).

At the 165 min the 1 km AGL updraft of M1500-2A, 2B, RP45, and RM45 was 13, 12, 12 and 12 m s$^{-1}$, respectively (Table 2). The downdrafts at this level were 16, 15, 14 and 11 m s$^{-1}$. M1500-2A and 2B low-level updrafts weaken considerably as they propagate towards the terrain. M1500-RP45 simulation updraft strengthens while the M1500-RM45 stays nearly the same (rounding). The midlevel updraft weakens in M1500-2A and 2B, intensifies in M1500-RP45 case, and remains about the same in M1500-RM45. The vorticity varies quite differently in these simulations as M1500-RM45 weakened as it approached the terrain then intensified slightly. M1500-2A and 2B weakened slightly as it propagated over the terrain. M1500-RP45 simulation intensified slightly as it approached the terrain then weakened significantly as it continued its track over the terrain.

In general, the terrain configuration of case 2A induced tornadogenesis in each of the three terrain heights indicating that this terrain geometry is the most likely to enhance vorticity along the gust front; Followed by the RP45, and the BSM of S14, which induced tornadogenesis in the cases with 500 and 1500 m terrain heights. Furthermore, the earlier tornadogenesis with higher terrain us to believe that there is no need to have a steep slope to enhance vorticity along the gust front, instead the approaching angle toward a terrain may allow the terrain blocking to change the direction of the inflow to enhance vorticity under the main updraft. This is also consistent with our finding that the cyclic nature is shortened in the simulations with increased terrain.

(c) Orographic Effects on Supercell Track

Using the method of S14 in which the storm location is identified using the updraft (at 1 km AGL) multiplied by the updraft helicity (1 – 6 km AGL) (UHW). They found that the track is shifted towards the north in their simulations with terrain to the left of the storm's motion, particularly for their 1500 m simulation.

Our investigation using modified terrain geometries has shown that M500 tracks are nearly identical to that of a simulation without terrain, although there are some timing differences but the track is basically the same. Each of the M1500 simulated storms is shifted to the south of the storm's motion and displaced the farthest in M1500-RM45 simulation, and that there is a generally rightward shift with respect to the storm's motion in the tracks as they propagate over the mountains. The general effect of elongating the terrain is to shift the track to left of storm motion early in M1000 and M1500 and toward the right when propagating up to and around the terrain. Although, these simulations were shifted southward this may be a result of a somewhat dissipative/weakening stage as convection is initiated near many of these storms.

These findings suggest that in general these terrain heights have a small to negligible effect on the track of supercells that are approaching the peak directly. This may also indicate that approaching the peak slightly to the south will produce a southward shift while the obverse would be true approaching slightly to the north. Another interpretation using the findings of Lin et al. (2005) is that our Vortex Froude number in several simulations is close to 1.5, where in their study was a transition point for continuous and discontinuous tracks of cyclones. This last interpretation is another area that needs further investigation as the time and length scales and maintenance mechanisms are very different between tropical cyclones, and supercells and tornadoes.

## 5. Concluding Remarks

The effects of elongated bell-shaped mountains, with different orientations and heights, on supercell thunderstorms were investigated in this study. The terrain produced gravity waves that modified the lee side of the terrain by generating alternating reductions and increases in the amounts of moisture, MLCAPE, and MLCIN. These gravity waves were strong enough to initiate convection in all of the 1000 m and 1500 m simulations. It is interesting to note that despite the RM45 orientation having the strongest blocking (in relation to the mean wind) it initiated convection later than M1000 and m1500-2A, 2B, and RP45. Convection associated with these storms was initiated from about the 90 – 180 min with the earliest initiated in the 2A followed by RP45, 2B, RM45 sequentially. The effects of reduced CAPE and increased CIN eventually would reduce the amount of energy available to the storm and was unable to sustain the storm.

The structure and development is nearly identical in all simulations out to the 75 min slight variations become insipient near the 90 min (about half way between the location of the initial warm bubble and the terrain peak). When comparing the M500 group there is the least variability between RM45, RP45, 2A, and 2B configurations (although there are differences between the simulations). The largest difference is between the 2A and other simulations where the updraft is considerably smaller and weaker at the 165 min, just after a period of very high rain fall. The 1000 m simulations exhibited variations in their intensity cycle that produced storms with similar structure at different times. Similar to the significate narrowing of the updraft in the 500 m 2A case there is a significant narrowing of updraft cores in the 1000 m RM45 and 2B cases after they undergo a similar period of very high rain fall. The 1500 m simulations behaved quite differently from the 500 and 1000 m cases experiencing extremely high rain fall rates and areal coverage, that in turn generated cold pools deep enough to initiate convection over the cold pool. The net effect of the additional convection was that the two storms started to "compete" with each other to ingest air and had a general weakening effect overall but help to produce more rain. Nearly all of the 1500 m simulations had a period where the rain rate approached 16 cm $h^{-1}$.

Our intensity investigation focused on the 1 and 5 km updrafts and the surface vorticity. The intensity of supercells was cyclic in all simulations, however, the period between intensity peaks were reduced in cases were more air was directed into the storm's inflow as compared to the no mountain (NMTN) control case. The increased (decreased) air flow created differences in the distributions of hydrometeors and increased (decreased) the rainfall rate and areal extent. This allowed the cold pool to be stronger in the simulations with more rain, most notably when the cold pools of the 1500 m storms became deep enough to initiate convection.

In addition, looking at these indicators of storm intensity we used a slightly stricter form of the National Weather Service's tornado detection algorithm to decide which terrain configuration would be most favorable for tornadogenesis. We found that despite being able to produce stronger blocking effects, the 1000 m mountains were generally the least favorable to tornadogenesis, with a tornado declared in only the M1000-2A case. The 1500 m mountains were most favorable for tornadogenesis, with a tornado declared in all cases except M1500-RM45. Surprisingly the RM45 configuration, the case with the strongest blocking in relation to the layer averaged mean wind, did not produce any tornadoes; this showed that it took more than blocking alone to generate a closed vortex throughout a depth of 1.5 km with sufficient winds to meet our criteria.

We have shown that additional air could modify supercellular convective life cycle through the redirection of additional air into the storms inflow. This additional air also increases the amount of rain in simulations with higher terrain deepening the cold pool (to the point that it may initiate convection in the 1000 m and 1500 m cases). This is of interest to now/forecasters as the increased rain can produce severe local flooding. Also of interest to now/forecasters is that terrain blocking effects alone are not sufficient enough to enhance tornadogenesis, but there has to be an increased likelihood of the terrain blocking effects to enhance the vorticity along the gust front and enhance the formation of a closed vortex beneath the main updraft. Furthermore, as high terrain produced a dramatic increase in vorticity this may lead to faulty attribution of damage to a tornado.

We investigated the proposed tracking method of S14 and found that with respect to modifying the orientation of elongated bell-shaped mountains using their UHW parameter for tracking a supercell thunderstorm was not robust. It was noted that this may be due to the impinging location slightly north or south could produce a northward or southward tendency for propagation. An alternate explanation offered that the Vortex Froude number was close to a transition region.

Areas where this study could be extended are incorporation of real terrain in these idealized simulations, such as configuring a domain that utilizes the terrain areas identified by Broyles and Crosbie (2004). Further, the flow regimes for these terrains using the WK82 sounding could be investigated; especially since some unexpected environmental evolutions formed (e.g. convection was initiated sooner in RP45 than RM45). This area still has much to be researched and additional attention is needed.

**Acknowledgments**

We would like to thank the National Oceanic and Atmospheric Administration (NOAA) Educational Partnership Program and the NOAA Earth System Research Laboratory for the use of their ZEUS supercomputer to conduct these simulations. Dr. George Bryan and NCAR are appreciated for allowing us to use CM1 and NCL, respectively. We thank IGES for the use of their Grid Analysis and Display System (GrADS) plotting software. This research was partially supported by the NOAA Cooperative Agreement No. NA06OAR4810187 and National Science Foundation Awards AGS-1265783, HRD-1036563, and OCI-1126543.

**References**
Bosart, L. F., Seimon, A., LaPenta, K. D., and Dickinson, M. J., 2006: Supercelltornadogenesis over complex terrain: The Great Barrington, Massachusetts, tornado on 29 May 1995. *Wea. Forec.*, 21, 897-922.
Bluestein, H. B., 2000: A tornadic supercell over elevated, complex terrain: The Divide, Colorado, storm of 12 July 1996. *Mon. Wea. Rev.*, 128, 795-809.
Broyles, J.C., and K.C. Crosbie*, 2004: Evidence of Smaller Tornado Alleys Across the United States Based on a Long Track F3-F5 Tornado Climatology Study from 1880-2003. Preprints, 22nd Conf. Severe Local Storms, Hyannis MA.
Bryan, G. and J. M. Fritsch, 2000: Moist absolute instability: The sixth static stability state. Bull. Amer. Meteor. Soc., 81, 1207-1230.
Bryan G. H., and Fritsch, J. M., 2002: A benchmark simulation for moist nonhydrostatic numerical models. Mon. Wea. Rev., 130, 2917–2928.
Chen, S.-H. and Lin, Y.-L., 2005: Effects of the basic wind speed and CAPE on flow regimes associated with a conditionally unstable flow over a mesoscale mountain. J. Atmos. Sci., 62, 331-350.
Ćurić et al., 2007, Ćurić, M., Janc, D., Vučković, V., 2007: Numerical simulation of a Cb cloud vorticity. Atmos. Res., 83, pp. 427–434
Deardorff, J.W., 1980: Stratocumulus-capped mixed layers derived from a threedimensional model. Bound.-Layer Meteor., 18, 495–527.
Dudhia, J., 1993: A nonhydrostatic version of the Penn State/NCAR mesoscale model: validation tests and simulation of an Atlantic cyclone and cold front. Mon. Wea. Rev. 121, 1493 – 1513.
Durran, D.R., Klemp, J.B., 1983: A compressible model for the simulation of moist mountain waves, Mon. Wea. Rev., 111, pp. 2341–2361
Emanuel, K.A., 1994, *Atmospheric Convection*, Oxford University Press, 580 pp.
Gal-Chen, T., and Somerville, R.C.J., 1975: On the use of a coordinate transformation for the solution of the Navier-Stokes equations. J. Comput. Phys., 17, 209-228, doi:10.1016/0021-9991(75)90037-6.
Grell, G.A., Dudhia, J., Stauffer, D.R., 1995. A description of the fifth-generation Penn State/NCAR mesoscale model (MM5), NCAR Tech. Note NCAR/TN-398 + STR.

Homar, V., 2003: Tornadoes over complex terrain: an analysis of the 28th August 1999 tornadic event in eastern Spain. *Atmos. Res.*, 67-68, 301-317.

LaPenta, K. D., Bosart, L.F., Galarneau, T. J., Deckinson, M. J., 2005: A Multiscale Examination of the 31 May 1998 Mechanicville, New York, Tornado. *Wea. Forec.*, 20, 494-516

Lin, Y.-L., Farley, R. D., and Orville, H. D., 1983: Bulk parameterization of the snow field in a cloud model. J. Clima. Appli. Meteor., 22, 1065-1092.

Lin, Y.-L., S.-Y. Chen, C. M. Hill, and C.-Y. Huang, 2005: Control parameters for tropical cyclones passing over mesoscale mountains, J. Atmos. Sci., 62, 1849-1866.

Lin, Y-L., 2007: *Mesoscale Dynamics*. Cambridge University Press, 630pp.

Markowski, P.M., Dotzek, N, 2011: A numerical study of the effects of orography on supercells. Atmospheric Research, 100, 457-478.

Morrison, H., Curry, J. A., and Khvorostyanov, V. I., 2005: A new doublemoment microphysics scheme for application in cloud and climate models. Part 1: Description. J. Atmos. Sci., 62, 1665–1677.

Schneider, D. G., 2009: The impact of terrain on three cases of tornadogenesis in the Great Tennessee Valley. *Elec. J. Oper. Meteor.*, EJ11, 33pp.

Smith, G. M., 2014 (S14): Numerical integration of orographic effects on supercell thunderstorms. Ph.D. Dissertation, North Carolina A&T State University, Greensboro, North Carolina, 105pp.

Thompson, R. L., Edwards, R., Mead, C. M., 2005: An update to the Supercell Composite and Significant Tornado Parameters. Preprints, *22nd Conf. Severe Local Storms*, Hyannis, MA.

Thompson, R. L., Edwards, R., Mead, C. M., 2007: Effective storm-relative helicity and bulk shear in supercell thunderstorm environments. Weather and Forecasting, 22, 102-115.

Weisman, M. L., and J. B. Klemp, 1982: The dependence of numerically simulated convective storms on vertical wind shear and buoyancy. Mon. Wea. Rev., 110, 504–520.

Weisman, M. L., and J. B. Klemp, 1984: The structure and classification of numerically simulated convective storms in directionally varying wind shears. Mon. Wea. Rev., 112, 2479–2498.

Wicker, L. J., and Skamarock, W. C., 2002: Time splitting methods for elastic models using forward time schemes. Mon. Wea. Rev., 130, 2088-2097

Xue, M., Droegemeier, K.K., Wong, V., Shapiro, A., Brewster, K., Carr, F., Weber, D., Lin, Y., Wang, D., 2001: The Advanced Regional Prediction System (ARPS)—a multi-scale nonhydrostatic atmospheric simulation and prediction model: Part II. Model physics and applications. Meteor. Atmos. Phys., 76, 143–165.

Table 1: Basic Unsaturated Moist Froude numbers, $F_w$, (second row) and Effective $F_w$ for indicated terrain configuration (remaining rows below second).

|       | 500 m | 1000 m | 1500 m |
|-------|-------|--------|--------|
| $F_w$ | 1.78  | 0.89   | 0.59   |
| RM45  | 1.91  | 0.95   | 0.63   |
| 2A    | 1.93  | 0.96   | 0.64   |
| 2B    | 2.76  | 1.38   | 0.91   |
| RP45  | 3.58  | 1.79   | 1.19   |

Table 2: Selected intensity parameters for the four terrain orientations and 3 heights.

|        | Time | 165 min |       |         | 180 min |       |         | 195 min |       |         |
|--------|------|---------|-------|---------|---------|-------|---------|---------|-------|---------|
|        | Case | w 1km   | w 5km | sfc vor | w 1km   | w 5km | sfc vor | w 1km   | w 5km | sfc vor |
| 500 m  | 2A   | 14.56   | 32.95 | 0.04    | 12.22   | 34.83 | 0.046   | 10.25   | 34.3  | 0.031   |
| 500 m  | 2B   | 14.01   | 35.83 | 0.037   | 12.35   | 36.89 | 0.03    | 8.44    | 32.94 | 0.016   |
| 500 m  | RP45 | 12.44   | 35.48 | 0.037   | 12.23   | 36.71 | 0.031   | 9.34    | 36.29 | 0.023   |
| 500 m  | RM45 | 11.73   | 39.49 | 0.019   | 10.84   | 35.28 | 0.058   | 11.65   | 36.08 | 0.052   |
|        |      |         |       |         |         |       |         |         |       |         |
| 1000 m | 2A   | 12.77   | 39.79 | 0.039   | 15.09   | 38.29 | 0.041   | 13.45   | 40.83 | 0.032   |
| 1000 m | 2B   | 11.74   | 32.35 | 0.049   | 14.16   | 35.15 | 0.035   | 9.26    | 37.37 | 0.037   |
| 1000 m | RP45 | 12.18   | 34.33 | 0.02    | 11.99   | 33.28 | 0.023   | 10.56   | 36.43 | 0.018   |
| 1000 m | RM45 | 12.25   | 38.08 | 0.025   | 14.27   | 36.22 | 0.061   | 12.25   | 28.64 | 0.034   |
|        |      |         |       |         |         |       |         |         |       |         |
| 1500 m | 2A   | 20.28   | 30.86 | 0.038   | 14.28   | 27.07 | 0.037   | 13.3    | 32.46 | 0.036   |
| 1500 m | 2B   | 18.27   | 32.91 | 0.037   | 11.64   | 29.61 | 0.033   | 11.91   | 31.24 | 0.028   |
| 1500 m | RP45 | 12.89   | 31.77 | 0.045   | 16.91   | 34.92 | 0.048   | 10.04   | 29.18 | 0.026   |
| 1500 m | RM45 | 14.48   | 33.38 | 0.026   | 14.51   | 32.82 | 0.02    | 14.5    | 35.15 | 0.025   |

Table 3: Indicates if the supercell thunderstorm met the criteria of the modified tornado detection algorithm to be declared tornadic. (For reference the BSM storms from S14 are included).

|       | BSM   | 2A    | 2B    | RM45 | RP45  |
|-------|-------|-------|-------|------|-------|
| M500  | Y 180 | Y 180 | N     | N    | Y 180 |
| M1000 | N     | Y 180 | N     | N    | N     |
| M1500 | Y 165 | Y 165 | Y 180 | N    | Y 165 |

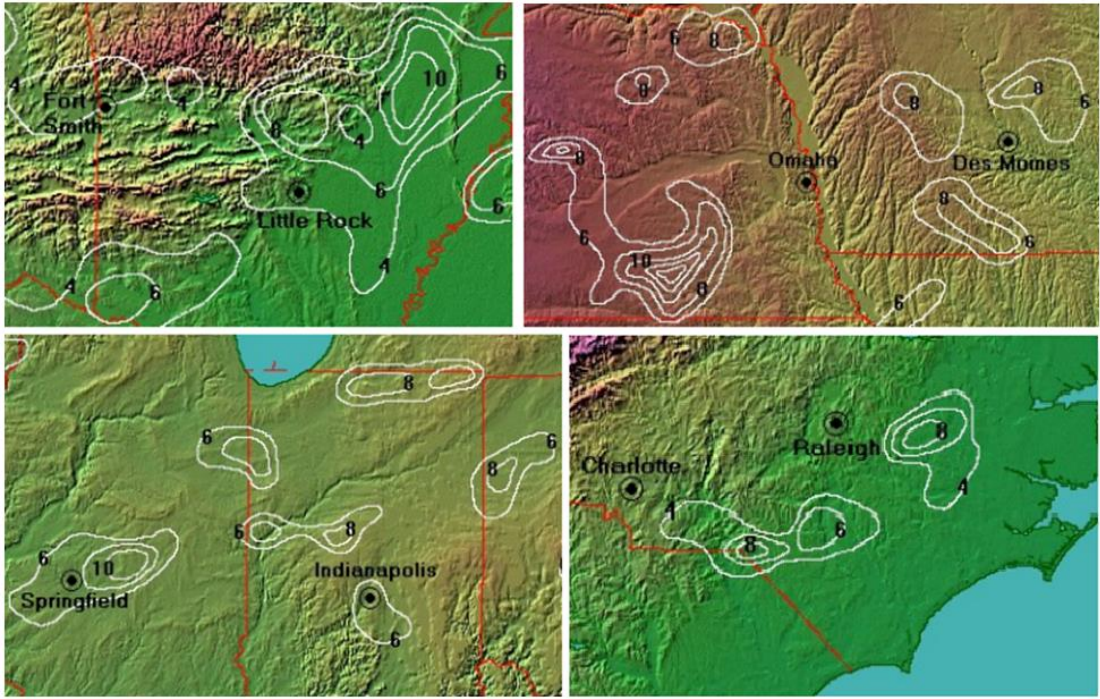

Figure 1: 1880-2003 F3-F5 Long Track Climatology of Violent Tornadoes. Adapted from Broyles and Crosbie (2004)

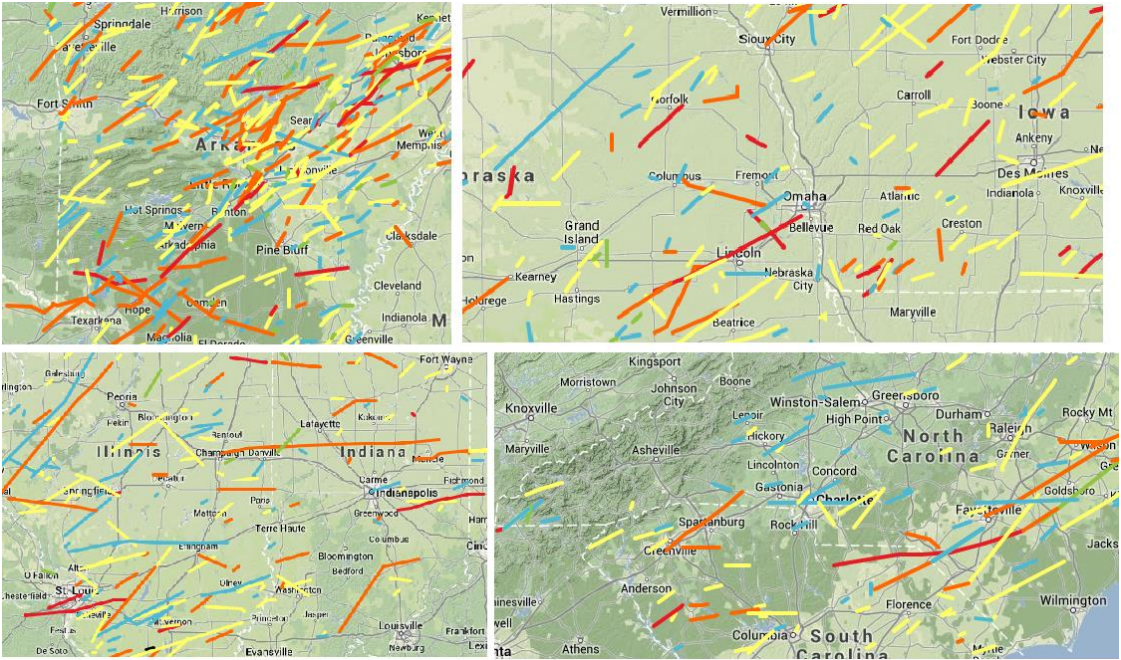

Figure 2: Tracks for all tornado occurrences in geographic sub-regions.

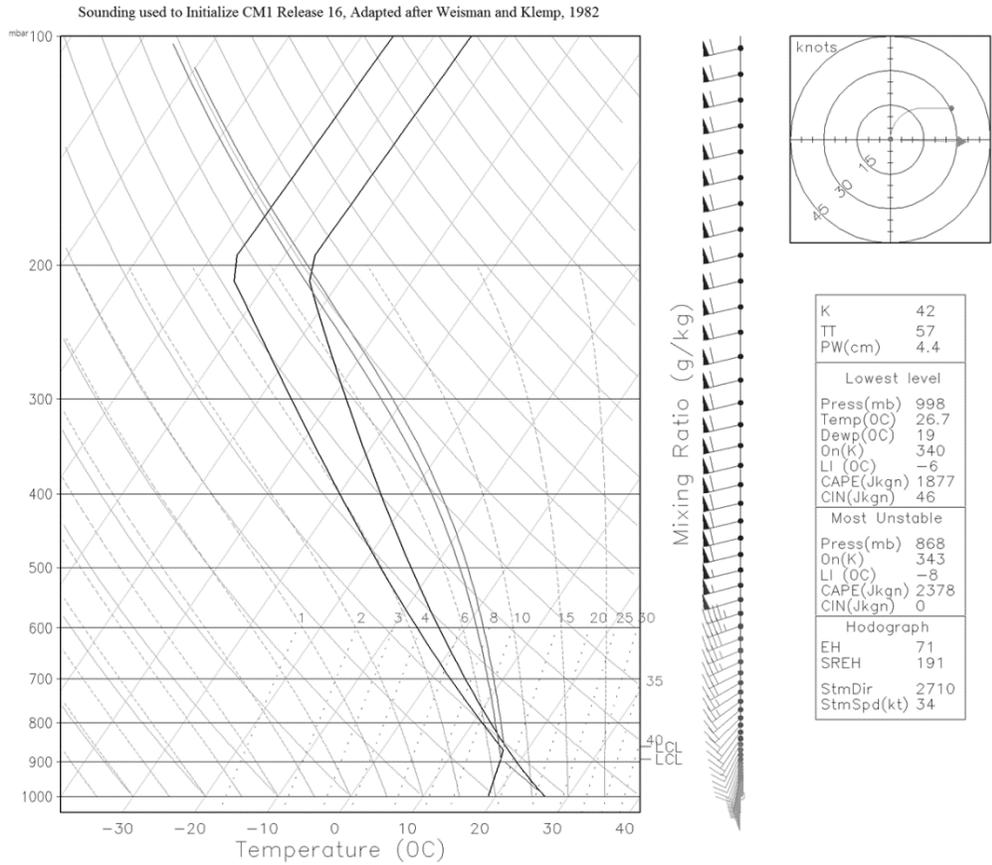

Figure 3: The sounding and wind profile used to initialize simulations in this study (Adapted after Weisman and Klemp, 1982).

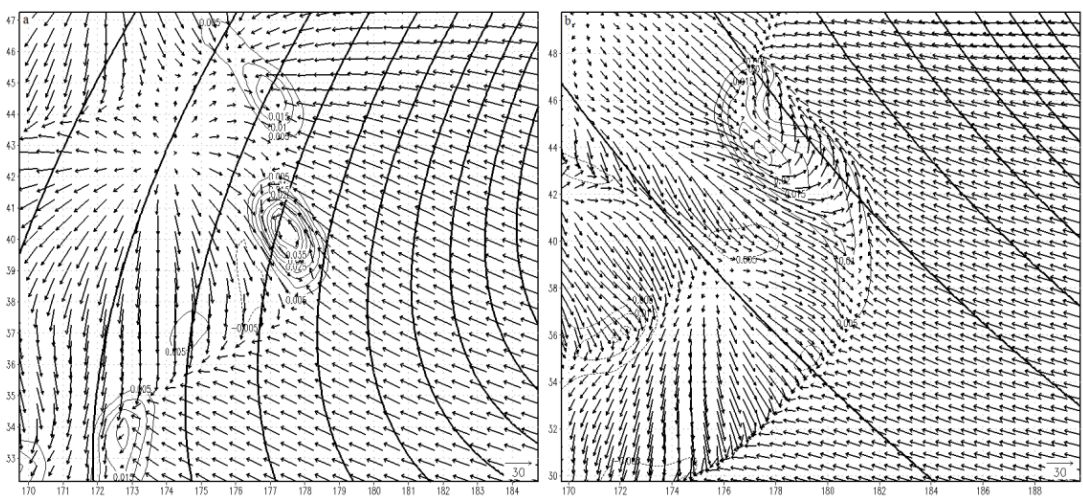

Figure 4: Example of a surface closed vortex that meets criteria 1 (a) and cross vortex sheer that does not meet criteria 1 (b)

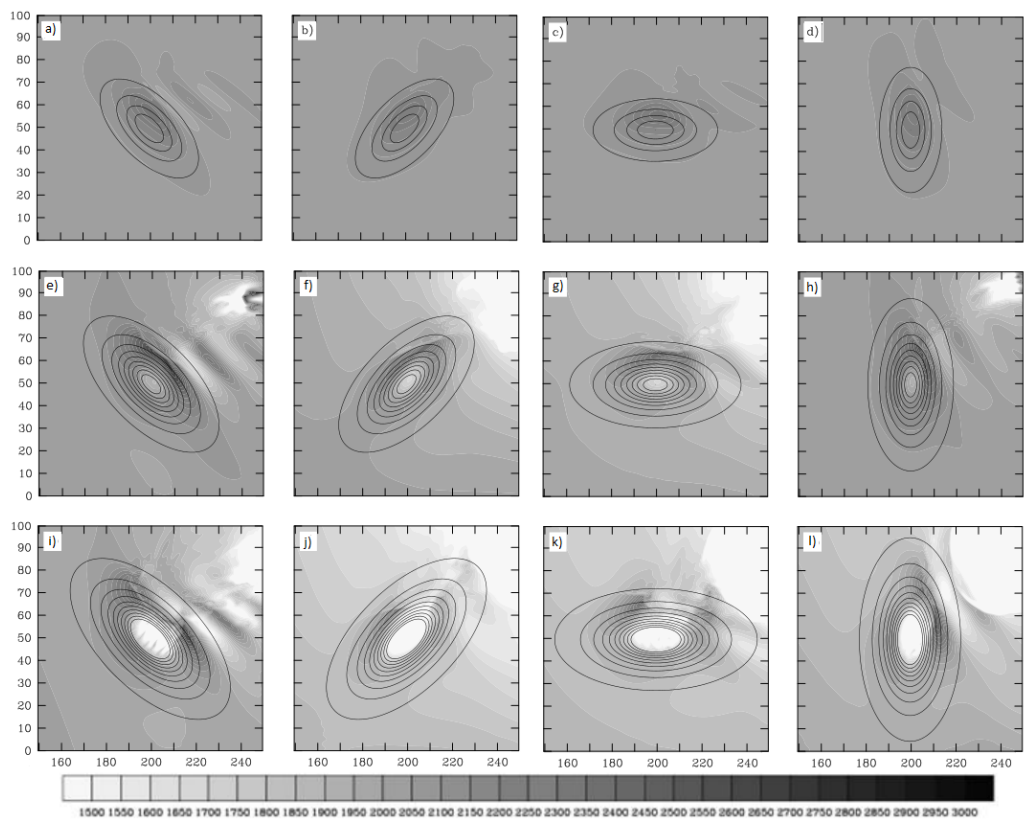

Figure 5: MLCAPE at the simulations third hour for the three varied heights (rows) 500, 1000, and 1500 m from top to bottom and the four different geometries (columns) RM45, RP45, 2A, and 2B from left to right.

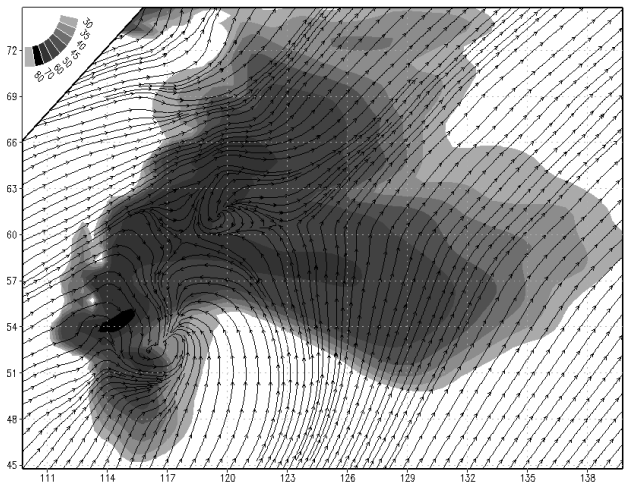

Figure 6: Reflectivity and wind stream-lines for the no-terrain control simulation (NMTN) at the 105 min. Note that the midlevel rotation is aligned with the updraft.

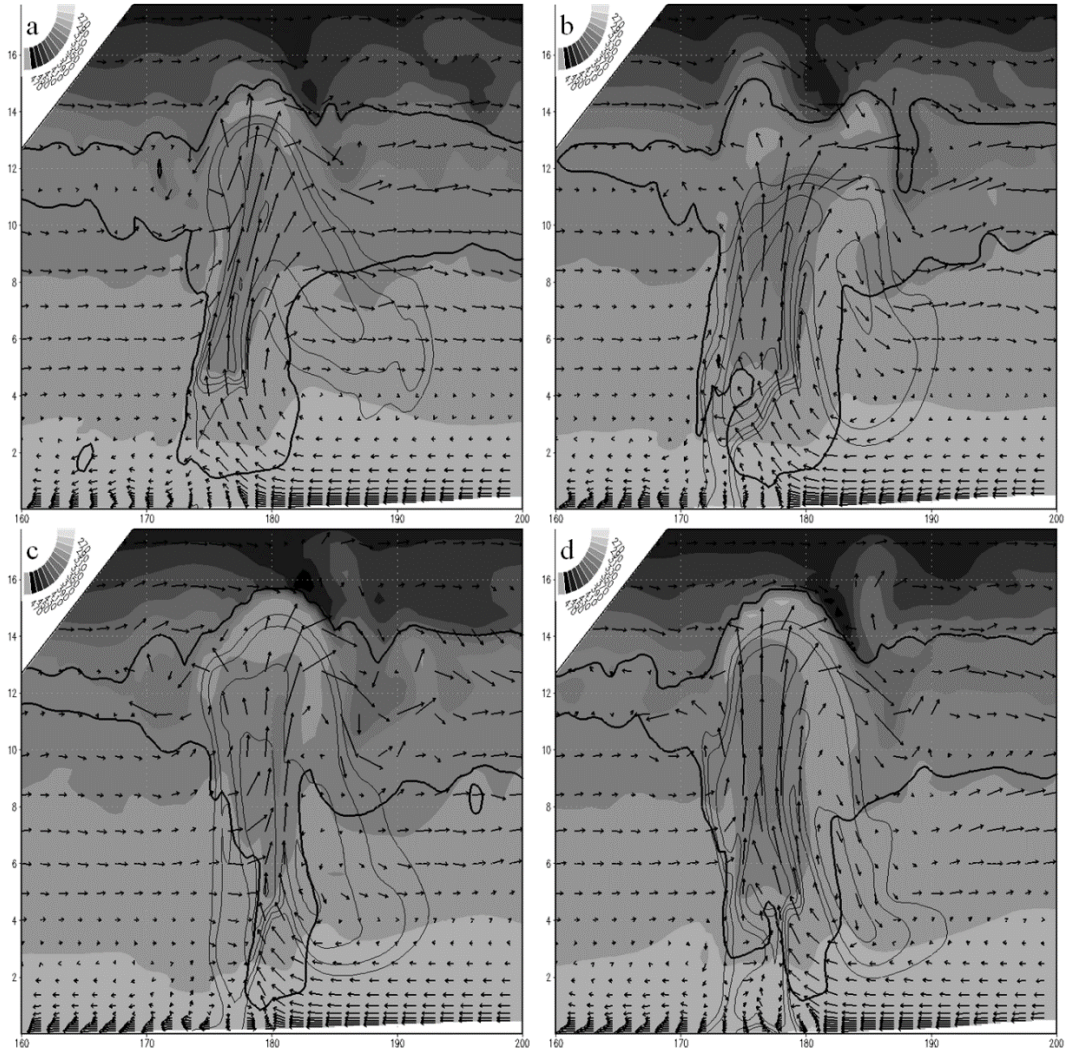

Figure 7: Zonal cross section of theta (shadeing), reflectivity (thin contour), cloud outline (thick contour), and wind vectors, at the 165 min for 500 m mountains and are a) RM45, b) RP45, c) 2A, d) 2B. Reflectivity values start at 50 dBZ and are contoured every 5 dBZ. The Cloud outline is the 0.5 g kg$^{-1}$ cloud and ice mixing ratios. The reference vector is in d and is the same for all panels. Cross section is along the direction of propagation (east-west) and is at the point of maximum UHW.

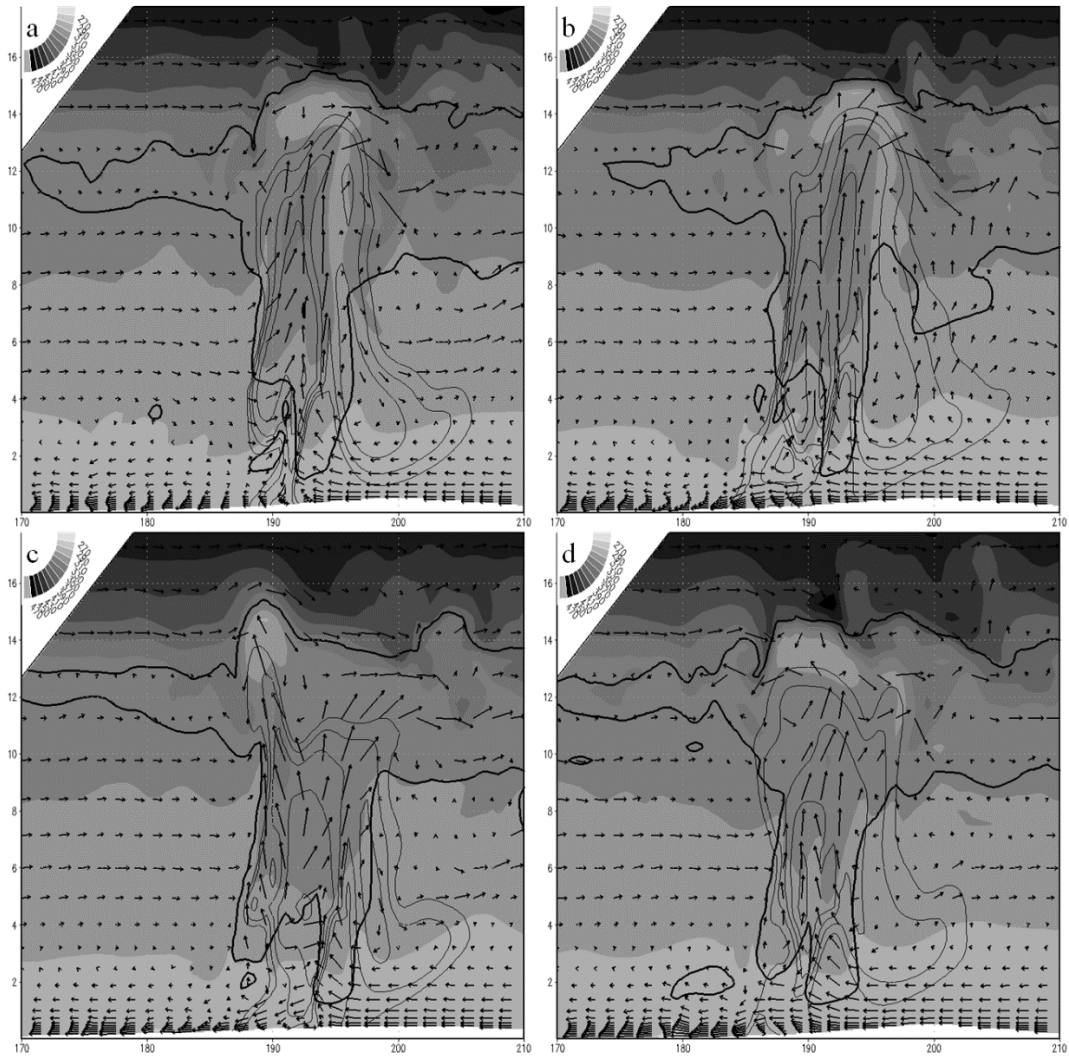

Figure 8: As in Figure 7, but at the 180 min.

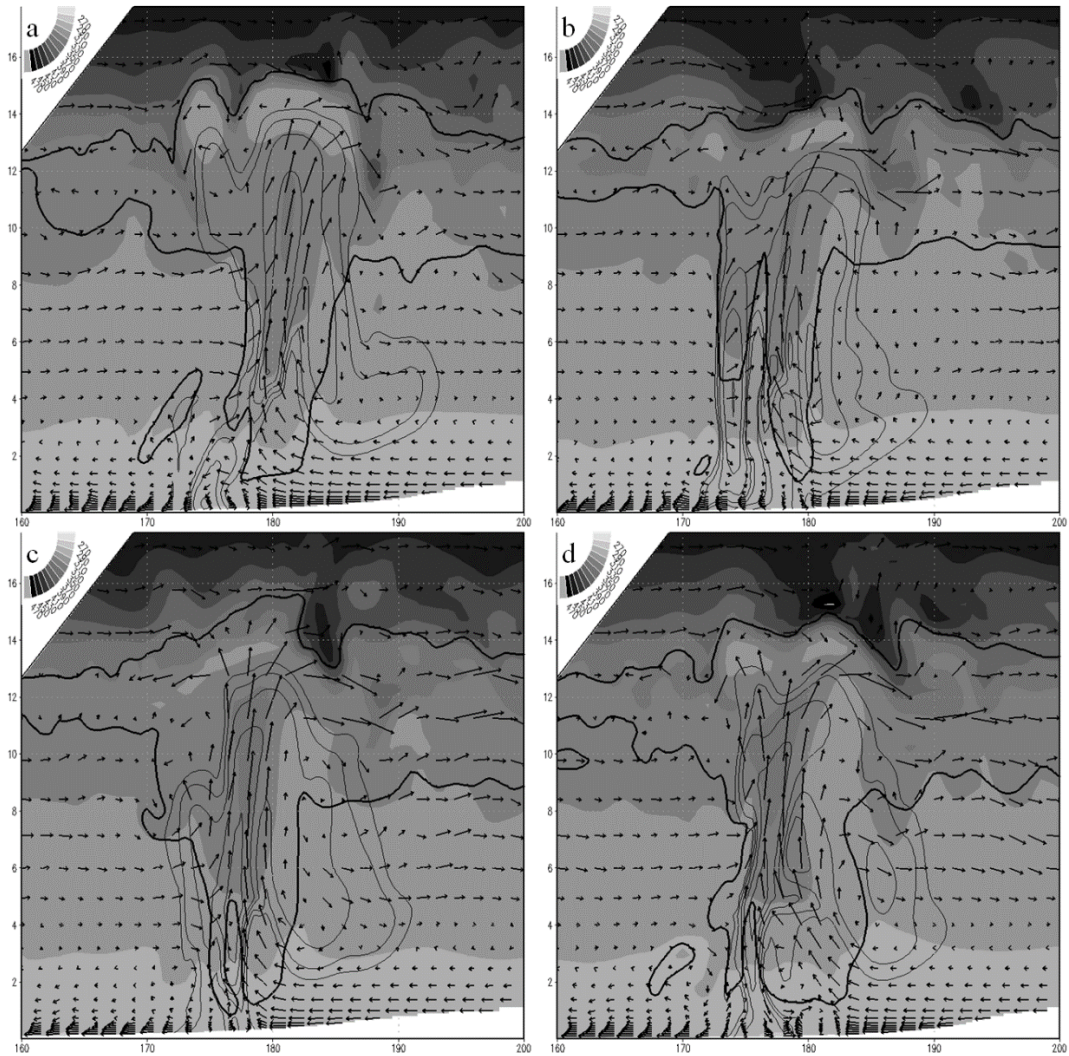

Figure 9: As in Figure 7, but for the 1000 m mountains.

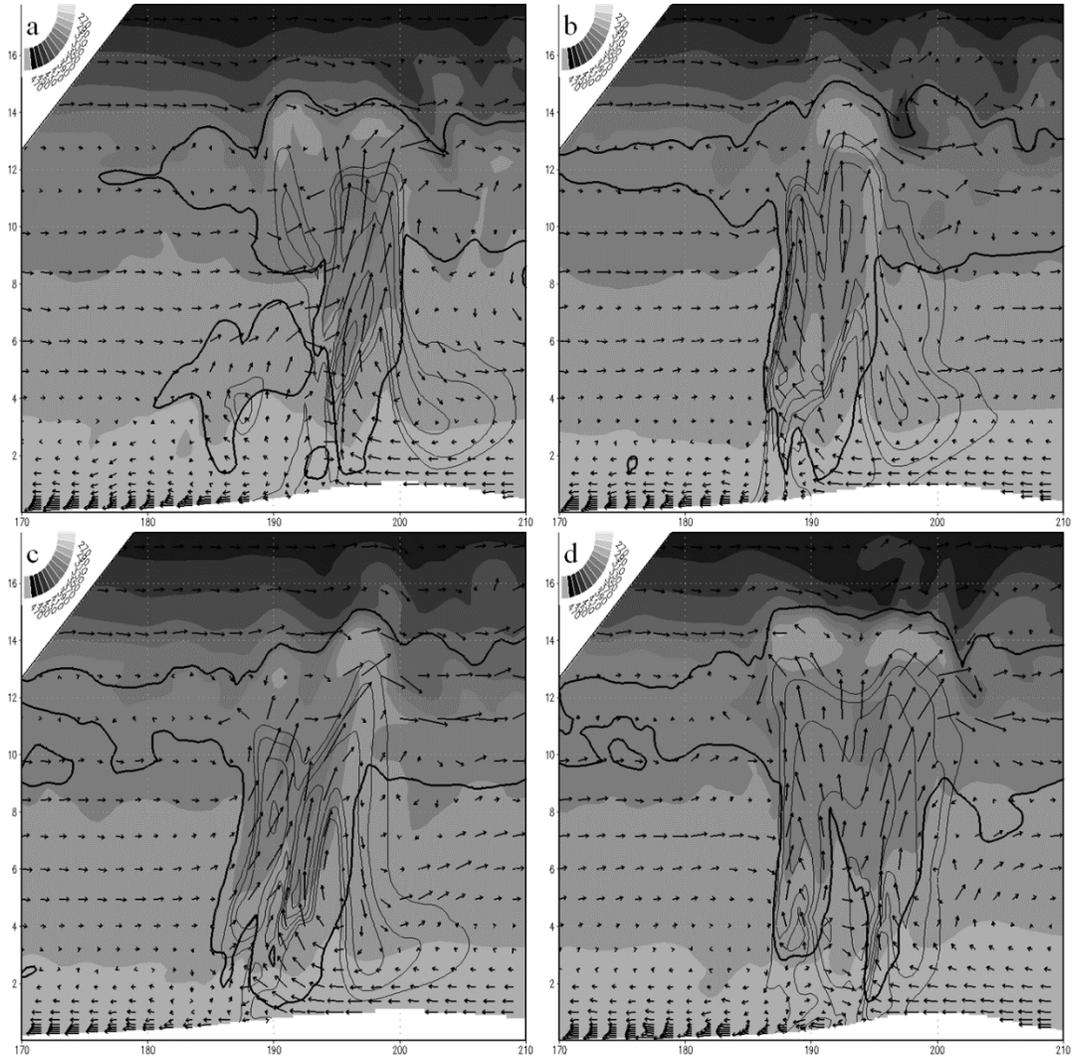

Figure 10: As in Figure 8, but for the 1000 m mountains.

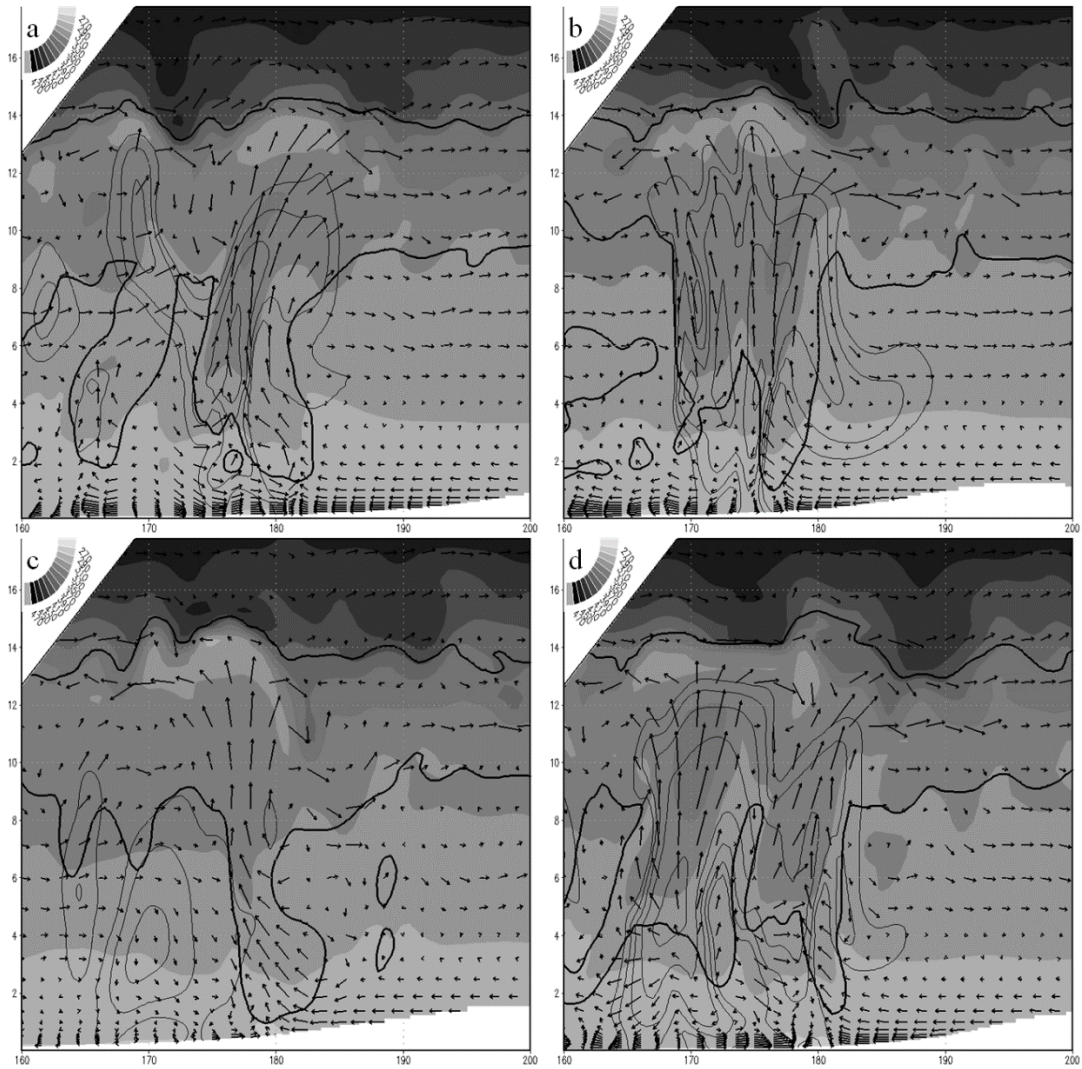

Figure 11: As in Figure 7, but for the 1500 m mountains.

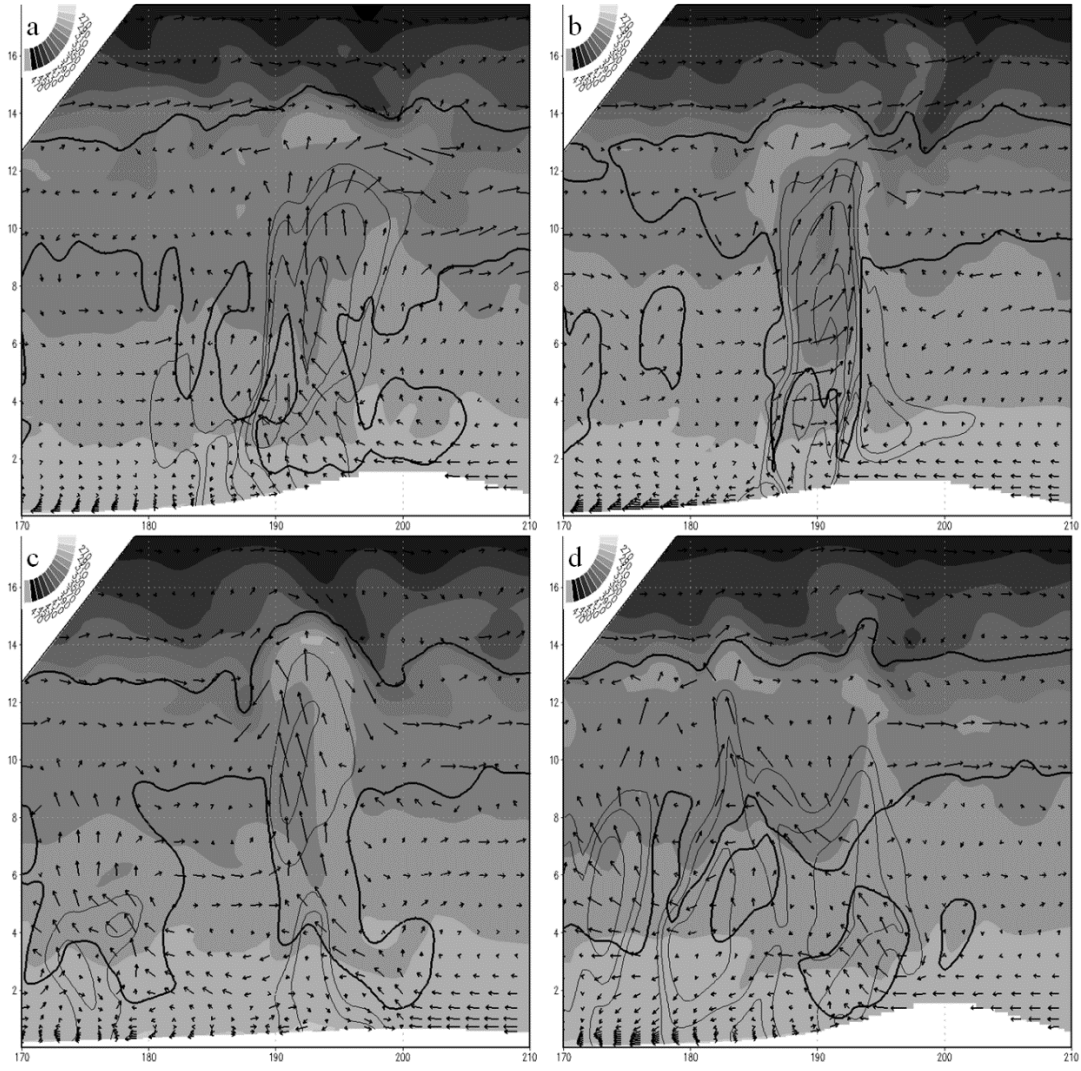

Figure 12: As in Figure 8, but for the 1500 m mountains.

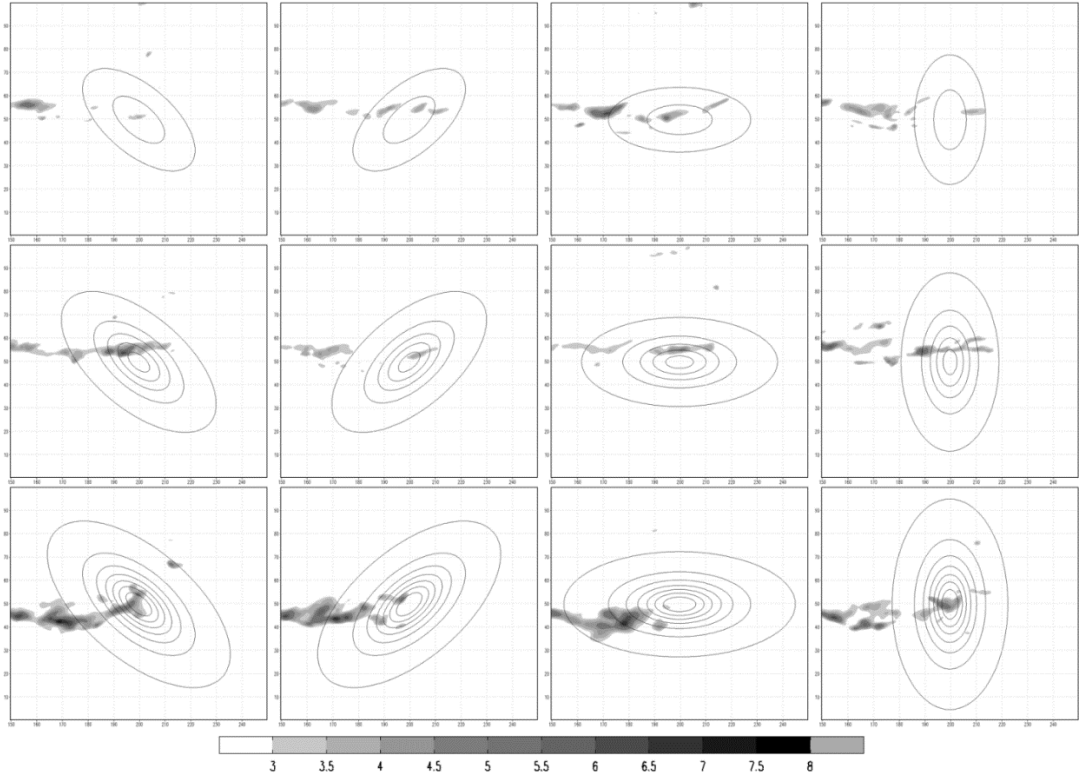

Figure 13: Total accumulated rain out to 210 min. Rows from top to bottom 500, 1000, 1500 m terrain heights. Columns from left to right RM45, RP45, 2A, 2B terrain orientations. Shading starts at 3 cm. Terrain contours start at 100 m and are every 200 m.

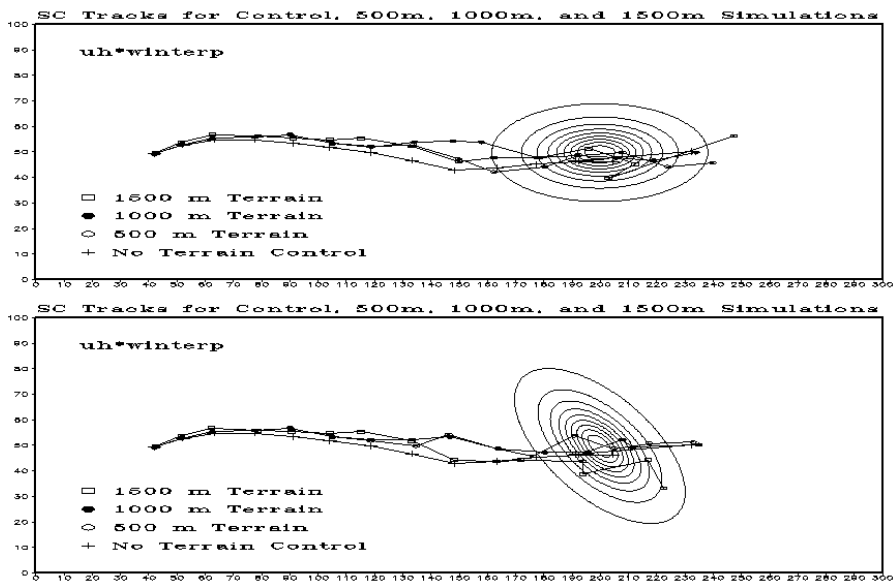

Figure 14: Tracks for a) the 1500 m 2A simulation and b) the 1500 m RM45 simulation.

23